\documentclass[prb, twocolumn]{revtex4}

\begin{document}
\title{Dynamics of collapsing and exploding Bose-Einstein condensates}
\author{Elizabeth~A. Donley,  Neil~R. Claussen, Simon~L. Cornish, Jacob~L.
  Roberts, Eric~A. Cornell\cite{byline} and Carl~E. Wieman}
\affiliation{JILA,
  National Institute of Standards and Technology and the University of
  Colorado, and the Department of Physics, University of Colorado,
  Boulder, Colorado 80309-0440}
\date{\today}

\begin{abstract}
We explored the dynamics of how a Bose-Einstein condensate
collapses and subsequently explodes when the balance of forces
governing the size and shape of the condensate is suddenly
altered. A condensate's equilibrium size and shape is strongly
affected by the inter-atomic interactions. Our ability to induce a
collapse by switching the interactions from repulsive to
attractive by tuning an externally-applied magnetic field yields a
wealth of detailed information on the violent collapse process. We
observe anisotropic atom bursts that explode from the condensate,
atoms leaving the condensate in undetected forms, spikes appearing
in the condensate wave function, and oscillating remnant
condensates that survive the collapse. These all have curious
dependencies on time, the strength of the interaction, and the
number of condensate atoms. Although ours would seem to be a
simple well-characterized system, our measurements reveal many
interesting phenomena that challenge theoretical models.
\end{abstract}

\maketitle \vspace{1 cm}

Although the density of the atoms in an atomic Bose-Einstein
condensate (BEC) is typically five orders of magnitude lower than
the density of air, the inter-atomic interactions greatly affect a
wide variety of BEC properties. These include static properties
like the BEC size and shape and the condensate stability, and
dynamic properties like the collective excitation spectrum and
soliton and vortex behavior. Since all of these properties are
sensitive to the inter-atomic interactions, they can be quite
dramatically affected by tuning the interaction strength and sign.

The vast majority of BEC physics is well described by mean-field
theory\cite{Dalfovo1999a}, in which the strength of the
interactions depends on the atom density and on one additional
parameter called the $s$-wave scattering length $a$. $a$ is
determined by the atomic species. When $a
> 0$, the interactions are repulsive. In contrast, when $a < 0$
the interactions are attractive and a BEC tends to contract to
minimize its overall energy. In a harmonic trap, the contraction
competes with the kinetic zero-point energy, which tends to spread
out the condensate. For a strong enough attractive interaction,
there is not enough kinetic energy to stabilize the BEC and it is
expected to implode. A BEC can avoid implosion only as long as the
number of atoms $N_0$ is less than a critical value given
by\cite{Ruprecht1995a}
\begin{equation}
\label{Eq:Stability} N_{cr} = k a_{ho}/|a| \;,
\end{equation}
where dimensionless constant $k$ is called the stability
coefficient. The precise value of $k$ depends on the aspect ratio
of the magnetic trap\cite{Gammal2001a}. $a_{ho}$ is the harmonic
oscillator length, which sets the size of the condensate in the
ideal-gas ($a=0$) limit.

Under most circumstances, $a$ is insensitive to external fields.
This is different in the vicinity of a so-called Feshbach
resonance, where $a$ can be tuned over a huge range by adjusting
the externally applied magnetic
field\cite{Tiesinga1992a,Tiesinga1993a}. This has been
demonstrated in recent years with cold $^{85}$Rb and Cs
atoms\cite{Roberts1998a,Courteille1998a,Vuletic1999a}, and with Na
and $^{85}$Rb Bose-Einstein
condensates\cite{Inouye1998a,Cornish2000a}. For $^{85}$Rb atoms,
$a$ is usually negative, but a Feshbach resonance at $\sim$155~G
allows us to tune $a$ by orders of magnitude and even change its
sign. This gives us the ability to create stable $^{85}$Rb
Bose-Einstein condensates\cite{Cornish2000a} and adjust the
inter-atomic interactions. We recently used this flexibility to
verify the functional form of equation (\ref{Eq:Stability}) and to
measure the stability coefficient to be
$k=0.46(6)$\cite{Roberts2001a}.

In this article, we study the dynamical response (``the
collapse'') of an initially stable BEC to a sudden shift of the
scattering length to a value more negative than the critical value
$a_{cr} = -ka_{ho}/N_{0 }$. We have observed many features of the
surprisingly complex collapse process, including the energies and
energy anisotropies of atoms that burst from the condensate, the
time scale for the onset of this burst, the rates for losing
atoms, spikes in the wave function that form during collapse,  and
the size of the remnant BEC that survives the collapse. The
unprecedented level of control provided by tuning $a$ has allowed
us to see how all of these quantities depend on the magnitude of
$a$, the initial number and density of condensate atoms, and the
initial spatial size and shape of the BEC before the transition to
instability.

A great deal of theoretical interest\cite{Kagan1997a,Kagan1998a,
Ueda1999b, Saito2001a, Saito2001b, Duine2001a} was generated by
BEC experiments in $^7$Li \cite{Bradley1997a}, for which the
scattering length is also negative and collapse events are also
observed\cite{Sackett1999a,Gerton2001a}. The $^7$Li experiments do
not employ a Feshbach resonance, so $a$ is fixed. This restricts
experimentation to the regime where the initial number of
condensate atoms is less than or equal to $N_{cr}$, and the
collapse is driven by a stochastic process. In addition, studies
of collapse dynamics in $^7$Li are complicated by a large thermal
component. Our ability to tune the scattering length, and most
notably explore the regime where the initial condensate is
``pure'' (near $T=0$) and the number $N_0$ is much larger than
$N_{cr}$, allows us to explore the dynamics and compare it with
theory in far more detail.

{\bf Experimental techniques.} The procedure for producing stable
$^{85}$Rb condensates has been described in detail
elsewhere\cite{Cornish2000a}. A standard double magneto-optical
trap (MOT) system\cite{Myatt1996a} was used to collect a cold
sample of $^{85}\rm{Rb}$ atoms in a low-pressure chamber. Once
sufficient atoms had accumulated in the low-pressure MOT, the
atoms were loaded into a cylindrically symmetric cigar-shaped
magnetic trap with trap the frequencies $\nu_{radial} = 17.5$~Hz
and $\nu_{axial} = 6.8$~Hz. Radio-frequency evaporation was then
used to cool the sample to $\sim$3~nK to form pure condensates
containing $>$90\% of the sample atoms. The final stages of
evaporation were performed at 162~G where the scattering length is
positive and stable condensates of up to 15,000 atoms could be
formed. After evaporative cooling, the magnetic field was ramped
adiabatically to 166~G (except where noted), where $a=0$. This
provided a well-defined initial condition with the BEC taking on
the size and shape of the harmonic oscillator ground state.

We could then adjust the mean-field interactions within the BEC to
a variety of values on time scales as short as 0.1~ms. The obvious
manipulation was to jump to some value of $a < a_{cr}$ to trigger
a collapse, but the tunability of $a$ also greatly aided in
imaging the sample. Usually the condensate size was below the
resolution limit of our imaging system (7$\mu$m FWHM). However, we
could ramp the scattering length to large positive values and use
the repulsive inter-atomic interactions to expand the BEC before
imaging, thus obtaining information on the pre-expansion
condensate shape and number.  A typical $a(t)$ sequence is shown
in Fig. \ref{Fig:Nvst}a. We have used a variety of such sequences
to explore many aspects of the collapse and enhance the visibility
of particular components of the sample.

{\bf{Condensate contraction and atom loss.}} When  the scattering
length is jumped to a value $a_{collapse} < a_{cr}$, a
condensate's kinetic energy no longer provides a sufficient
barrier against collapse. As described in
ref.~\onlinecite{Kagan1998a}, during collapse one might expect a
BEC to contract until losses from density-dependent inelastic
collisions\cite{Roberts2000a} would effectively stop the
contraction.  This contraction would roughly take place on the
time scale of a trap oscillation, and the density would sharply
increase after $T_{rad}/4 \simeq 14$~ms, where $T_{rad}$ is the
radial trap period. How does this picture compare to what we have
actually seen?

A plot of the condensate number $N$ vs $\tau_{evolve}$ for
$a_{collapse} = -30~a_0$ and $a_{init} = +7~a_0$ is presented in
Fig. \ref{Fig:Nvst}b. $N$ was constant for some time after the
jump until atom loss suddenly began at $t_{collapse}$. After the
jump the condensate was smaller than our resolution limit, so we
could not observe the contraction directly. But we observed that
the post-expansion condensate widths changed very little with time
$\tau_{evolve}$ before $t_{collapse}$. From this we infer that the
bulk BEC did not contract dramatically before loss began. We
modelled the contraction by putting the $a(t)$ sequence that we
used for the experiment into the equations of ref.
\onlinecite{Perez-Garcia1997a} for the approximate evolution of
the BEC shape. The model predicted that by $t_{collapse}$, the
radial and axial widths had contracted by $\sim$20\% and
$\sim$4\%, respectively, from their initial values. This
contraction only corresponds to a 50\% increase in the average
density to $2.5 \times 10^{13}/{\rm cm}^3$. Using the decay
constants from ref. \onlinecite{Roberts2000a}, this density gives
an atom loss rate, $\tau_{decay}$, that is far smaller than what
we observe and does not have the observed sudden onset.

For the data in Fig. \ref{Fig:Nvst}b and most other data presented
below, we jumped to $a_{quench}=0$ in $0.1$~ms after a time
$\tau_{evolve}$ at $a_{collapse}$. We believe that the loss
immediately stopped after the jump. This interpretation is based
on the surprising observation that the quantitative details of
curves such as that shown in Fig. \ref{Fig:Nvst}b did not depend
on whether the collapse was terminated by a jump to $a_{quench} =
0$ or $a_{quench} = 250~a_0$.

We have measured loss curves like that in Fig. \ref{Fig:Nvst}b for
many different values of $a_{collapse}$. The collapse time versus
$a_{collapse}$ for $N_0 = 6000$ presented in Fig.
\ref{Fig:Collapse_time} shows the strong dependence on
$a_{collapse}$. Reducing the initial density by a factor of $\sim
4$ (with a corresponding increase in volume) by setting $a_{init}
= + 89~a_0$ for one value of $a_{collapse}$ (-15 $a_0$), increased
$t_{collapse}$ by $\sim 3$.

The atom loss time constant $\tau_{decay}$ depended only weakly
on $a_{collapse}$ and $N_0$. For the range of $a_{collapse}$ shown
in Fig. \ref{Fig:Collapse_time}, $\tau_{decay}$ did not depend on
$a_{collapse}$ or $N_0$ outside of the experimental noise
($\sim$20\%). On average, $\tau_{decay}$ was 2.8(1)~ms. For the
very negative value of $a_{collapse} \simeq - 250~a _0$, however,
$\tau_{decay}$ did decrease to 1.8~ms for $N_0 = 6,000$ and 1.2~ms
for $N_0 = 15,000$.

{\bf{Burst atoms.}} As indicated by Fig. \ref{Fig:Nvst}b, atoms
leave the BEC during the collapse. There are at least two
components to the expelled atoms. One component (the ``missing
atoms'') is not detected. The other component emerges as a burst
of detectable spin-polarized atoms with energies much greater than
the initial condensate's energy but much less than the magnetic
trap depth. The burst-energy dependencies on $a_{collapse}$ and
$N_0$ are complex, but since they will provide a stringent test of
collapse theories, we present them in detail.

The angular kinetic-energy distribution with which the burst atoms
are  expelled from the condensate can most accurately be measured
by observing their harmonic oscillations in the trap, as
illustrated in Fig. \ref{Fig:Burst}a. For example, half of a
radial period after the expulsion ($T_{rad}/2$), all atoms return
to their initial radial positions. Well before or well after this
``radial focus'', the burst cloud is too dilute to be observed.
Fortunately, at the radial focus, oscillations along the axial
trap axis are near their outer turning points and the axial energy
can be found from the length of the stripe of atoms along the
axial axis. The radial energy can be found with the same procedure
for an axial focus. The sharpness of the focus also provides
information on the time extent of the burst.

Figure \ref{Fig:Burst}b shows an image of a radial focus. The size
scales for the burst focus and the remnant were well separated
since the latter was not expanded before imaging. Figure
\ref{Fig:Burst}c shows cross sections of the burst focus and fits
to the burst and the thermal cloud. The burst energy distributions
were well fit by Gaussians characterized by a temperature that was
usually different for the two trap directions. The burst energy
fluctuated from shot to shot by up to a factor of 2 for a given
$a_{collapse}$. This variation is far larger than the measurement
uncertainty or the variation in initial number (both $\sim$ 10\%),
and its source is unknown. (We also discuss observed structural
variability when we present the jet measurements below.)

Although the burst energies varied from shot to shot, the average
value was well-defined and showed trends far larger than the
variation. The axial and radial burst energies versus
$a_{collapse}$ are shown in Fig. \ref{Fig:Energies}a and
\ref{Fig:Energies}b for $N_0 = 6,000$ and 15,000, respectively.
The burst-energy anisotropy shown in Fig. \ref{Fig:Energies}c
depended on $N_{0}$, $a_{collapse}$, and $a_{init}$.

When we interrupted the collapse with a jump back to $a = 0$ as
discussed above, we also interrupted the growth of the burst.  The
``interrupted'' burst atoms still refocused after sitting at $a =
0$ for the requisite half trap period.  The energy of the atoms in
the interrupted bursts appeared to be the same, but the number of
atoms was smaller.  By changing the time at which the collapse was
interrupted we could measure the time dependence of the creation
of burst atoms.  For the conditions of Fig. \ref{Fig:Nvst}b, the
number of burst atoms $N_{burst}$ grew with $\tau_{evolve}$ with a
time constant of 1.2~ms starting at 3.5~ms and reaching an
asymptotic final number of $\sim$2500 for all times $\geq$7~ms.
$N_{burst}$ varied randomly by $\sim$20\% for the data in Fig.
\ref{Fig:Energies}, but on average the fraction of atoms going
into the burst was about 20\% of $N_0$ and did not depend on
$a_{collapse}$ or $N_0$.

{\bf{Remnant Condensate.}} After a collapse, a ``remnant''
condensate containing a fraction of the atoms survived with nearly
constant number for more than 1 second and oscillated in a highly
excited collective state with the two lowest modes with $\nu
\simeq 2\nu_{axial}$ and $\nu \simeq 2\nu_{radial}$ being
predominantly excited. (The measured frequencies were $\nu =
13.6(6)$~Hz and $\nu = 33.4(3)$~Hz.) To find the oscillation
frequencies, the widths of the condensate were measured as a
function of time spent at $a_{collapse}$.

The number of atoms in the remnant depended on $a_{collapse}$ and
$N_0$, and in general was not limited by the critical number,
$N_{cr}$. The stability condition in equation (\ref{Eq:Stability})
determined the collapse point but it did not constrain
$N_{remnant}$. A fixed fraction of $N_0$ went into the remnant
independent of $N_0$, so that smaller condensates often ended up
with $N_{remnant} < N_{cr}$, but larger condensates rarely did.
The fraction of atoms that went into the remnant decreased with
$|a_{collapse}|$, and was $\sim$40\% for $|a_{collapse}| < 10~a_0$
and $\sim$50\% for $|a_{collapse}| > 100~a_0$. We do not think
that the surface-wave excitations\cite{Pattanayak2001a} are
responsible for stabilizing the remnant since we excite
large-amplitude breathing modes. For $N_0 = 6,000$ and
$|a_{collapse}| < 10~a_0$, more atoms were lost than the number
required to lower $N_{remnant}$ to below $N_{cr}$.

Since $N_{burst}$ was independent of $a_{collapse}$ but
$N_{remnant}$ decreased with $|a_{collapse}|$, the number of
missing atoms increased with $|a_{collapse}|$. Interestingly, the
number of missing atoms also increased with $N_0$, but the
fraction of missing atoms versus $a_{collapse}$ was equal for $N_0
= 6,000$ and $N_0 = 15,000$ and was $\sim$40\% for $|a_{collapse}|
< 10~a_0$ and $\sim$70\% for $|a_{collapse}| \ge 100~a_0$. The
missing atoms were presumably either expelled from the condensate
at such high energies that we could not detect them ($> 20~\mu$K),
or they were transferred to untrapped atomic states or undetected
molecules.

{\bf Jet formation.} Under very specific experimental conditions,
we observed streams of atoms with highly anisotropic velocities
emerging from the collapsing condensates. These ``jets'' are
distinguished from the ``burst'' in that the jets have much lower
kinetic energy (on the order of a few nanokelvin), in that their
velocity is nearly purely radial, and in that they appear only
when the collapse is interrupted (i.e., by jumping to $a_{quench}
= 0$) during the period of number loss. Collapse processes that
were allowed to evolve to completion (until $N \simeq
N_{remnant}$) were not observed to emit jets. Examples are shown
in Fig. \ref{Fig:Jets} for different $\tau_{evolve}$ for the
conditions of Fig. \ref{Fig:Nvst}b. The jet size and shape varied
from image to image even when all conditions were unchanged, and
as many as three jets were sometimes observed to be emitted from
the collapse of a single condensate. The jets were also not always
symmetric about the condensate axis.

We believe that these jets are manifestations of local ``spikes''
in the condensate density that form during the collapse and expand
when the balance of forces is changed by quenching the collapse.
We can estimate the size of the spikes using the uncertainty
principle. After a jump to $a_{quench}=0$, the kinetic energy of
the atoms in the resulting jet is equal to the confinement energy
that the spike had prior to quenching the collapse, i.e.,
$\frac{1}{2}mv^2 = \frac{\hbar^2}{4m\sigma^2}$, where $\sigma$ is
the width of the spike in the wave function. The anisotropy of the
jets indicates that the spikes from which they originated were
also highly anisotropic, being narrower in the radial direction.
From the widths and the number of atoms in the jets, we can
estimate the density in the spikes. Plots of the number of jet
atoms and the inferred density in the spikes versus
$\tau_{evolve}$ are presented in Fig. \ref{Fig:Jets2}. The jets
exhibited variability in energy and number that was larger than
the $\sim$~10\% measurement noise.

{\bf{Overview of the current theoretical models.}} Several
theoretical
papers\cite{Kagan1997a,Kagan1998a,Ueda1999b,Saito2001a,Saito2001b,Duine2001a}
have considered the problem of collapse of a BEC when the number
of atoms exceeds the critical number.  These treatments all use a
mean-field approach and describe the condensate dynamics using the
Gross-Pitaevskii (GP) equation. In most
cases\cite{Kagan1997a,Kagan1998a,Saito2001a,Saito2001b}, the loss
mechanism is three body recombination, but Duine and
Stoof\cite{Duine2001a} propose the loss arises from a new elastic
scattering process.  In both cases the loss is density dependent
and so the loss rate is quite sensitive to the dynamics of the
shape of the condensate. Since a full three-dimensional
anisotropic time-dependent solution to the GP equation is very
difficult, these calculations have used various approximations to
calculate the time evolution of the condensate shape. Kagan,
Muryshev, and Shlyapnikov\cite{Kagan1998a} numerically integrate
the GP equation for the case of an isotropic trap and large values
of the three body recombination coefficient. In this regime, the
condensate smoothly contracts in a single, collective collapse.
Saito and Ueda\cite{Saito2001a,Saito2001b} perform a similar
numerical solution for the isotropic case but with smaller values
of $K_3$ and observe localized spikes to form in the wave function
during collapse. Duine and Stoof\cite{Duine2001a} model the
dynamics for the anisotropic case, but use a Gaussian
approximation rather than an exact numerical solution.

These calculations have all been carried out over a certain range
of parameters, but none have been done for the specific range of
parameters that correspond to our experimental situations. None of
the predictions in these papers match our measurements except for
the general feature that atoms are lost from the condensate.  Also
we see several phenomena that are not discussed in these papers.
Whether this lack of agreement is due to the fact that these
calculations do not scale to our experimental situation or do not
contain the proper physics remains to be seen.

{\bf{Theoretical Challenges.}} Collapsing $^{85}$Rb condensates
present a rather simple system with quite dramatic behavior. This
behavior might provide a rigorous test of mean-field theory when
it is applied to our experimental conditions. Some of our
particularly puzzling results are:
\begin{itemize}
\item
The decay constant $\tau_{decay}$ is independent of both $N_0$ and
$|a_{collapse}|$ for $|a_{collapse}| < 100~a_0$, and only weakly
depends on these quantities for larger $|a_{collapse}|$.
\item
The burst energy per atom dramatically increases with initial
condensate number.
\item
The number of burst atoms is constant versus $a_{collapse}$.
\item
The number of cold remnant BEC atoms surviving the violent
collapse varies between much less and much more than $N_{cr}$,
depending on $N_0$ and $a_{collapse}$, but the fraction of remnant
atoms, burst atoms, and missing atoms are independent of $N_0$.
\end{itemize}

{\bf{Outlook.}} From the experimental point of view there remain
questions to be answered. For one, is the burst coherent? It may
be possible to answer this by generating a sequence of ``half
bursts'' and see if they interfere. For another, where do the
missing atoms go? If molecules and/or relatively high-energy atoms
are being created, can we detect them?

It is clear that adjustable interactions opens up a fascinating
new regime for BEC studies.

\vspace{0.5 cm} \noindent {\bf{ACKNOWLEDGMENTS.}} We extend our
thanks to Sarah Thompson for laboratory assistance and to Stephan
D\"{u}rr, Gora Shlyapnikov, Henk Stoof, Murray Holland, Masahito
Ueda, and Rembert Duine for helpful discussions. This research has
been supported by the ONR, NSF, ARO--MURI, and NIST. S.~L.~C.
acknowledges the support of a Lindemann Fellowship.

 \begin{figure}[p]
 \caption{
A example of a ramp applied to the scattering length, and a plot
of the condensate number versus time after a jump to a negative
scattering length. a. A typical $a(t)$ sequence. $a_0 =
0.529~{\rm{\AA}}$ is the Bohr radius. The scattering length is
jumped at $t = 0$ in 0.1~ms from $a_{init}$ to $a_{collapse}$,
where the BEC evolves for a time $\tau_{evolve}$. The field is
carefully controlled so that magnetic-field noise translates into
fluctuations in $a_{collapse}$ on the order of $\sim 0.1~a_0$ in
magnitude. The collapse is then interrupted with a jump to
$a_{quench}$, and the field is ramped in 5~ms to a large positive
scattering length which makes the BEC expand. After 7.5~ms of
additional expansion, the trap is turned off in 0.1~ms and 1.8~ms
later the density distribution is probed using destructive
absorption imaging with a $40~\mu$s laser pulse (indicated by the
vertical bar). The increase in $a$ from $a_{collapse}$ to
$a_{expand}$ is far too rapid to allow for the BEC to expand
adiabatically. On the contrary, the smaller the BEC before
expansion, the larger the cloud at the moment of imaging. Thus we
can readily infer the relative size of the bulk of the BEC just
prior to the jump to $a_{quench}$. The density of the expanded BEC
is so low that the rapid transit of the Feshbach resonance
pole\cite{Stenger1999c} during the trap turn-off and the
subsequent time spent at $B=0$ ($a=-400~a_0$) both have a
negligible effect. b. The number of atoms remaining in the BEC
versus $\tau_{evolve}$ at $a_{collapse}=-30~a_0$. We observed a
delayed and abrupt onset of loss. The solid line is a fit to an
exponential with a best-fit value of $t_{collapse} \simeq
3.7(5)$~ms for the delay.}
 \label{Fig:Nvst}
 \end{figure}
\begin{figure}
 \caption{
The collapse time versus $a_{collapse}$ for 6000-atom condensates.
The vertical line indicates  $a_{cr}$ for $N_0 = 6,000$. The data
were acquired with $a_{init} = a_{quench} = 0$ (to within $\sim
2~a_0$).
 }
 \label{Fig:Collapse_time}
 \end{figure}
  \begin{figure}
 \caption{
A burst focus. a.  Conceptual illustration of a radial burst
focus. b. An image of a radial burst focus   taken 33.5~ms after a
jump from $a_{init}=0$ to -30~$a_0$ for $N_0 = 15,000$.
$T_{rad}/2$ = 28.6~ms, which indicates that the burst occurred
4.9(5)~ms after the jump. The axial energy distribution for this
burst corresponded to an effective temperature of 62~nK. The image
is $60 \times 310~\mu$m. c. Radially averaged cross section of b
with a Gaussian fit to the burst energy distribution. The central
$100~\mu$m were excluded from the fit to avoid distortion in the
fit due to the condensate remnant ($\sigma = 9 \mu$m) and the
thermal cloud ($\sigma = 17 \mu$m). The latter is present in the
pre-collapse sample due to the finite temperature and appears to
be unaffected by the collapse. The dashed line indicates the fit
to this initial thermal component. Note the offset between the
centers of the burst and the remnant. This offset varies from shot
to shot by an amount comparable to the offset shown. }
 \label{Fig:Burst}
 \end{figure}
  \begin{figure}
 \caption{
Burst energies and energy anisotropies. a. and b. The axial and
radial burst energies versus $a_{collapse}$ for $N_0 \simeq 6,000$
and $N_0 \simeq 15,000$, respectively.  On average, ten burst
focuses were measured for each trap direction at each value of
$a_{collapse}$ studied. The energies were higher for the
larger-$N_0$ condensates over the full range of $a_{collapse}$
studied. The vertical and horizontal error bars indicate the
standard error of the measurements and the uncertainty in
$a_{collapse}$ arising from the magnetic-field calibration,
respectively. For several of the points, the uncertainties are
smaller than the symbol size and the error bars are not visible.
c. The ratio of the radial to the axial energies, which is a
measure of the burst anisotropy. For values of $|a_{collapse}|$
just past $a_{cr}$, the burst was isotropic for both $N_0 \simeq
6,000$ and $N_0 \simeq 15,000$. At larger values of
$|a_{collapse}|$, larger-$N_0$ condensates gave rise to stronger
anisotropies. For $N_0 = 6000$, $\lambda$ was 1.6, and for $N_0$ =
15,000, $\lambda$ was 2. When instead we started at $a_{init} =
+100~a_0$, the BEC was initially more anisotropic ($\lambda =
2.4$), but the burst became more isotropic, with $E_{ax}$ going up
by $\sim$40\% and $E_{rad}$ dropping by $\sim$60\% for $N_0 =
15,000$ and $a_{collapse} = -100~a_0$.}
 \label{Fig:Energies}
 \end{figure}

 \begin{figure}
 \caption{
Jet images for a series of $\tau_{evolve}$ values for the
conditions of Fig. \ref{Fig:Nvst}b. The evolution times were 2, 3,
4, 6, 8, and 10~ms (from a to f). Each image is $150 \times
255~\mu$m. The bar indicates the optical depth scale. An expansion
to $a_{expand}=+250~a_0$ was applied, so the jets are longer than
for the quantitative measurements explained in the text. The jets
were longest (i.e., most energetic) and contained the most atoms
at values of $\tau_{evolve}$ for which the slope of the loss curve
(Fig. \ref{Fig:Nvst}b) was greatest. A tiny jet is barely visible
for $\tau_{evolve} \simeq 2$~ms (image a), which is 1.7~ms before
$t_{collapse}$. The images also show how the number of condensate
atoms decreases with time. The time from the application of
$a_{quench}$ until the acquisition of the images was fixed at
5.2~ms. }
 \label{Fig:Jets}
 \end{figure}
 \begin{figure}
 \caption{
Quantitative jet measurements. a. The number of atoms in the jets
versus $\tau_{evolve}$ for the conditions of Fig. \ref{Fig:Nvst}b.
b. The spike density inferred from the kinetic energies of the
jets. The bars indicate the range of shot-to-shot variability. For
the analysis, we assumed the jets were disk shaped since the
magnetic trap is axially symmetric. The images were taken
perpendicular to the axial trap axis, viewing the disks edge-on.
The jets expanded with $v \simeq 1$~mm/s, which corresponds to a
kinetic energy of $\sim$6~nK and a radial pre-quench Gaussian rms
width of $\sim 0.5~\mu$m. Since the axial size was below our
resolution limit, we could not measure the axial expansion rate.
For estimating the spike density, we assumed an axial width equal
to the harmonic oscillator length. The atom density in the spikes
decreased for larger values of $|a_{collapse}|$, and was half as
large for $a_{collapse} = -100~a_0$ as for $-30~a_0$.}
 \label{Fig:Jets2}
 \end{figure}

\begin{thebibliography}{10}

\bibitem[*]{byline} Quantum Physics Division, National Institute of Standards and Technology.

\bibitem{Dalfovo1999a}
Dalfovo, F., Giorgini, S., Pitaevskii, L.~P., \& Stringari, S.
\newblock Theory of {B}ose-{E}instein condensation in trapped gases.
\newblock {\em Rev. Mod. Phys.\/} {\bf 71}, 463--512 (1999).

\bibitem{Ruprecht1995a}
Ruprecht, P.~A., Holland, M.~J., Burnett, K., \& Edwards, M.
\newblock Time-dependent solution of the nonlinear {S}chr{\"o}dinger equation
  for {B}ose-condensed trapped neutral atoms.
\newblock {\em Phys. Rev. A\/} {\bf 51}, 4704--4711 (1995).

\bibitem{Gammal2001a}
Gammal, A., Frederico, T., \& Tomio, L.
\newblock Critical number of atoms for attractive Bose-Einstein condensates with cylindrically
symmetrical traps.
\newblock {\em Phys. Rev. A\/} (submitted.) (2001).

\bibitem{Tiesinga1992a}
Tiesinga, E., Moerdijk, A.~J., Verhaar, B.~J., \& Stoof, H.~T.~C.
\newblock Conditions for Bose-Einstein condensation in magnetically trapped atomic cesium.
\newblock {\em Phys. Rev. A\/} {\bf 46}, R1167 (1992).

\bibitem{Tiesinga1993a}
Tiesinga, E., Verhaar, B.~J., \& Stoof, H.~T.~C.
\newblock Threshold and resonance phenomena in ultracold ground-state
collisions.
\newblock {\em Phys. Rev. A\/} {\bf 47}, 4114--4122 (1993).

\bibitem{Courteille1998a}
Courteille, Ph., Freeland, R.~S., Heinzen, D.~J., van Abeelen,
F.~A., \& Verhaar, B.~J.
\newblock Observation of a Feshbach resonance in cold atom scattering.
\newblock {\em Phys. Rev. Lett.\/} {\bf 81}, 69--72 (1998).

\bibitem{Roberts1998a}
Roberts, J.~L., {\em et~al.\/}
\newblock Resonant magnetic field control of elastic scattering in cold $^{85}$Rb.
\newblock {\em Phys. Rev. Lett.\/} {\bf 81}, 5109--5112 (1998).

\bibitem{Vuletic1999a}
Vuleti\'c, V., Kerman, A.~J., Chin, C., \& Chu, S.
\newblock Observation of low-field Feshbach resonances in collisions of cesium atoms.
\newblock {\em Phys. Rev. Lett.\/} {\bf 82}, 1406--1409 (1999).

\bibitem{Inouye1998a}
Inouye, S. {\em et~al.\/}
\newblock Observation of Feshbach resonances in a Bose-Einstein condensate.
\newblock {\em Nature\/} {\bf 392}, 151--154 (1998).

\bibitem{Cornish2000a}
Cornish, S.~L., Claussen, N.~R., Roberts, J.~L., Cornell, E.~A.,
\& Wieman,  C.~E.
\newblock Stable $^{85}${R}b {B}ose-{E}instein condensates with widely tunable
  interactions.
\newblock {\em Phys. Rev. Lett.\/} {\bf 85}, 1795--1798 (2000).

\bibitem{Roberts2001a}
Roberts, J.~L., {\em et~al\/}.
\newblock Controlled collapse of a Bose-Einstein condensate.
\newblock {\em Phys. Rev. Lett.\/} {\bf 86}, 4211--4214 (2001).

\bibitem{Kagan1997a}
Kagan, Y., Surkov, E.~L., \& Shlyapnikov, G.~V.
\newblock Evolution and global collapse of trapped Bose condensates under variations
of the scattering length.
\newblock {\em Phys. Rev. Lett.\/} {\bf 79}, 2604--2607 (1997).

\bibitem{Kagan1998a}
Kagan, Y., Muryshev, A.~E., \& Shlyapnikov, G.~V.
\newblock Collapse and {B}ose-{E}instein condensation in a trapped {B}ose Gas
  with negative scattering length.
\newblock {\em Phys. Rev. Lett.\/} {\bf 81}, 933--937 (1998).

\bibitem{Ueda1999b}
Ueda, M. \& Huang, K.
\newblock Fate of a {B}ose-{E}instein condensate with an attractive
  interaction.
\newblock {\em Phys. Rev. A\/} {\bf 60}, 3317--3320 (1999).

\bibitem{Saito2001a}
Saito, H. \& Ueda, M.
\newblock Power laws and collapsing dynamics of a trapped Bose-Einstein
  condensate with attractive interactions.
\newblock {\em Phys. Rev. A\/} {\bf 63}, 043601--7 (2001).

\bibitem{Saito2001b}
Saito, H. \& Ueda, M.
\newblock Intermittent implosion and pattern formation of trapped Bose-Einstein
  condensates with an attractive interaction.
\newblock {\em Phys. Rev. Lett.\/} {\bf 86}, 1406--1409 (2001).

\bibitem{Duine2001a}
Duine, R.~A. \& Stoof, H. T.~C.
\newblock Explosion of a collapsing Bose-Einstein condensate.
\newblock {\em Phys. Rev. Lett.\/} {\bf 86}, 2204--2207 (2001).

\bibitem{Bradley1997a}
Bradley, C.~C., Sackett, C.~A. \& Hulet, R.~G.
\newblock Bose-Einstein condensation of Lithium: Observation of Limited condensate number.
\newblock {\em Phys. Rev. Lett.\/} {\bf 78}, 985--988 (1997).

\bibitem{Sackett1999a}
Sackett, C.~A., Gerton, J.~M., Welling, M., \& Hulet, R.~G.
\newblock Measurements of collective collapse in a Bose-Einstein condensate
  with attractive interactions.
\newblock {\em Phys. Rev. Lett.\/} {\bf 82}, 876--879 (1999).

\bibitem{Gerton2001a}
Gerton, J.~M., Strekalov, D., Prodan, I., \& Hulet, R.~G.
\newblock Direct observation of growth and collapse of a Bose-Einstein
  condensate with attractive interactions.
\newblock {\em Nature\/} {\bf 408}, 692--695 (2001).

\bibitem{Myatt1996a}
Myatt, C.~J., Newbury, N.~R., Ghrist, R.~W., Loutzenhiser, S., \&
Wieman, C.~E.
\newblock Multiply loaded magneto-optical trap.
\newblock {\em Opt. Lett.\/} {\bf 21}, 290--292 (1996).

\bibitem{Roberts2000a}
Roberts, J.~L., Claussen, N.~R., Cornish, S.~L., \& Wieman, C.~E.
\newblock Magnetic field dependence of ultracold inelastic collisions near a
  {F}eshbach resonance.
\newblock {\em Phys. Rev. Lett.\/} {\bf 85}, 728--731 (2000).

\bibitem{Stenger1999c}
Stenger, J., {\em et~al.\/}.
\newblock Strongly enhanced inelastic collisions in a {B}ose-{E}instein
  condensate near {F}eshbach resonances.
\newblock {\em Phys. Rev. Lett.\/} {\bf 82}, 2422--2425 (1999).

\bibitem{Perez-Garcia1997a}
P{\'e}rez-Garc{\'i}a, V.~M., Michinel, H., Cirac, J.~I.,
Lewenstein, M., \&
  Zoller, P.
\newblock Dynamics of {B}ose-{E}instein condensates: {V}ariational solutions of
  the {G}ross-{P}itaevskii equations.
\newblock {\em Phys. Rev. A\/} {\bf 56}, 1424--1432 (1997).

\bibitem{Pattanayak2001a}
Pattanayak, A.~K., Gammal, A., Sackett, C.~A., \& Hulet, R.~G.
\newblock Stabilizing an attractive Bose-Einstein condensate by driving a
  surface collective mode.
\newblock {\em Phys. Rev. A\/} {\bf 63}, 033604--4 (2001).

\end{thebibliography}
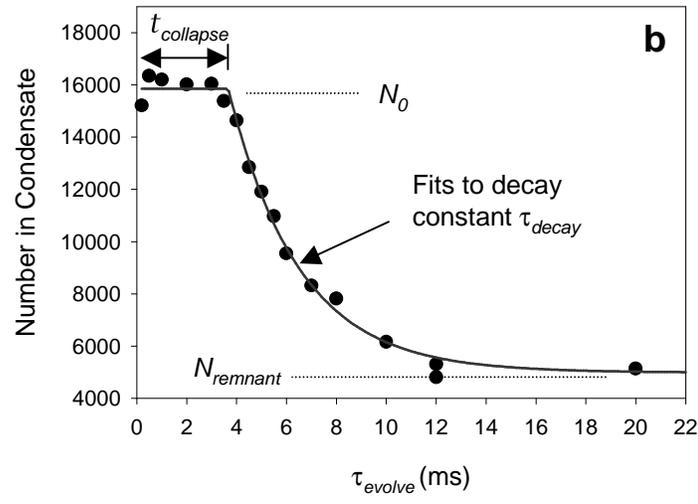
\end{document}